\documentclass[aps, twocolumn, superscriptaddress, nofootinbib]{revtex4-2}

\usepackage{amsmath}
\usepackage{amssymb}
\usepackage{mathrsfs}
\usepackage{xspace}
\usepackage{braket}
\usepackage{bbold}
\usepackage{graphicx}
\usepackage[dvipsnames]{xcolor}

\definecolor{fun}{rgb}{0.7, 0, 0.35}
\usepackage[colorlinks, hypertexnames=false, allcolors=fun]{hyperref}
\usepackage[capitalise]{cleveref} 
\crefname{section}{Sec.}{Secs.}
\Crefname{section}{Section}{Sections}
\renewcommand{\citep}[1]{Ref.~\cite{#1}}
\renewcommand{\citealp}[1]{Refs.~\cite{#1}}
\renewcommand{\autoref}[1]{Eqs.~\eqref{#1}}

\newcommand{\eg}{{e.g.,\/}\xspace}
\newcommand{\ie}{{i.e.,\/}\xspace}
\newcommand{\etal}{{\it et~al.\/}\xspace}

\newcommand{\pd}{\partial}

\newcommand{\mc}[1]{\mathcal{#1}}
\newcommand{\mcc}[1]{\mathfrak{#1}}
\newcommand{\msf}[1]{\mathsf{#1}}

\renewcommand{\vec}[1]{{\boldsymbol{#1}}}
\newcommand{\favr}[1]{\langle #1 \rangle}

\newcommand{\ee}{\mathrm{e}}
\newcommand{\ii}{\mathrm{i}}
\newcommand{\dd}{\mathrm{d}}

\renewcommand{\Re}{\text{Re}\,}
\renewcommand{\Im}{\text{Im}\,}

\newcommand{\spatialk}{\mathsf{k}}

\newcommand{\lie}{\mathrm{\text{\pounds}}}
\newcommand{\total}[1]{\msf{#1}}
\newcommand{\m}[1]{$\smash{#1}$}
\newcommand{\hmag}{\mathbb{h}}
\newcommand{\amag}{\mathbb{a}}
\newcommand{\hamp}{\mcc{h}}
\newcommand{\aamp}{\mcc{a}}

\begin{document}

\title{Self-consistent interaction of linear gravitational and electromagnetic waves in non-magnetized plasma}

\author{Deepen Garg}
\email{dgarg@princeton.edu}
\affiliation{Department of Astrophysical Sciences, Princeton University, Princeton, New Jersey 08544, USA}
\author{I.\ Y.\ Dodin}
\affiliation{Department of Astrophysical Sciences, Princeton University, Princeton, New Jersey 08544, USA}
\affiliation{Princeton Plasma Physics Laboratory, Princeton, NJ 08543, USA}

\date{\today}

\begin{abstract}
This paper explores the hybridization of linear metric perturbations with linear electromagnetic (EM) perturbations in non-magnetized plasma for a general background metric. The local wave properties are derived from first principles for inhomogeneous plasma, without assuming any symmetries of the background metric. First, we derive the effective (``oscillation-center'') Hamiltonian that governs the average dynamics of plasma particles in a prescribed quasimonochromatic wave that involves metric perturbations and EM fields simultaneously. Then, using this Hamiltonian, we derive the backreaction of plasma particles on the wave itself and obtain gauge-invariant equations that describe the resulting self-consistent gravito-electromagnetic (GEM) waves in a plasma. The transverse tensor modes of gravitational waves are found to have no interaction with the plasma and the EM modes in the geometrical-optics limit. However, for \textit{longitudinal} GEM modes with large values of the refraction index, the interplay between gravitational and EM interactions in plasma can have a strong effect. In particular, the dispersion relation of the Jeans mode is significantly affected by electrostatic interactions. As a spin-off, our calculation also provides an alternative resolution of the so-called Jeans swindle.
\end{abstract}

\maketitle

\section{Introduction}

Detection of correlated emission of gravitational waves (GWs) and electromagnetic (EM) radiation \cite{ref:abbott17d, ref:abbott19b, ref:soares17} has ushered a new era of multi-messenger astronomy. The idea behind this venture is to use observations of gravitational and EM signals synergistically to learn more about sources of these signals (which can range from compact object mergers \cite{ref:abbott17d, ref:abbott19b, ref:soares17} to early Universe \cite{ref:adshead21, ref:orlando22, ref:afzal23, tex:antoniadis23}) than each signal type allows individually. This synergy can be used even more efficiently if we better understand interactions between the multiple messengers, \ie GWs and EM radiation. Although some of these interactions have been studied (for example, the photon--graviton conversion \cite{ref:fujita20} and the EM fields acting as the source term for the gravitational tensor modes \cite{ref:caprini01, ref:caprini06}), the backreaction of the plasma is usually considered in the literature only to a limited degree or not at all \cite{ref:isliker06, ref:brodin00, ref:brodin10b, ref:brodin01, ref:brodin05, ref:macedo83, ref:servin01, ref:moortgat03, ref:forsberg10a, ref:bamba18, ref:mendonca02b}. Much like in the study of the interaction of GWs with matter \cite{ref:asseo76, ref:chesters73, ref:madore72, ref:madore73, ref:baym17,ref:flauger18, ref:asenjo20, ref:barta18, ref:duez05a}, the polarization of the metric perturbations is usually assumed to be transverse-traceless a~priori based on the symmetry considerations for an isotropic homogeneous background. Such tensor modes are independent of the scalar and vector perturbations \cite{book:mukhanov, ref:bardeen80, book:durrer20} in symmetric backgrounds, and they have indeed been shown to not interact significantly with EM perturbations \cite{ref:cruise83} and uniformly distributed neutral matter \cite{ref:flauger18, ref:baym17}. However, theory of the coupling of GWs with a generic distribution of matter remains to be developed.

On small enough scales, the distribution of matter can be nonuniform and the background may not be considered isotropic and homogeneous. Then, the gauge-invariant modes are not independent of each other \cite{ref:clarkson10, ref:pereira07}. As the background parameters evolve in an inhomogeneous background, the dispersion and the polarization of the waves evolve as well \cite{ref:bamba18}, and the distinction between the tensor modes and other collective oscillations such as the Jeans mode becomes blurred. This is not dissimilar to EM waves in plasmas, where transverse modes can gradually transform into longitudinal modes when the plasma parameters are inhomogeneous \cite[Chap.~13]{book:stix}. Studying such transformations (known as mode conversion in general wave theory \cite{book:tracy}) requires a systematic study in which all the perturbations are treated on an equal footing and the polarization is derived rather than inherited from the studies of GWs in highly symmetric backgrounds.

Here, we present such a theory for the first time, specifically, for non-magnetized plasma. We adopt Whitham's variational approach \cite{book:whitham, ref:dougherty70, ref:dewar77, my:amc, phd:ruiz17}, which has also been used for GWs in the past \cite{ref:isaacson68a, ref:maccallum73, ref:araujo89, ref:butcher09, my:spinhall, ref:andersson21} and facilitates tractable analytical calculations without assuming any GW properties a~priori (except linearity and the short-wavelength limit). One advantage of this approach is that it yields an asymptotic theory of geometrical-optics (GO) waves for a generic smooth background. An additional advantage is that Whitham's approach yields a gauge-invariant set of equations \cite{my:gwquasi}. This general gauge invariance can also be used then to check how accurate models are for specific dispersion operators, as detailed in \cref{sec:eulerem,sec:waveeqh}. Our formulation develops on the one presented in our \citep{my:gwjeans}, where we studied GWs in neutral gases. Here, we introduce EM field as another degree of freedom. First, we derive the effective (``oscillation-center'') Hamiltonian that governs the average dynamics of plasma particles in a prescribed quasimonochromatic wave that involves spacetime-metric perturbations and EM fields simultaneously. Then, using this Hamiltonian, we derive the backreaction of plasma on the wave itself and obtain gauge-invariant equations that describe the resulting self-consistent gravito-electromagnetic (GEM) waves in a plasma.

We find that any interaction between the gravitational modes and the EM modes are absent (within the limitations of our model) in the cold plasma limit, thus requiring thermal effects for any interaction. This is similar to the conclusions of \citep{ref:flauger18}, which reports that cold matter produces no significant effect on the dispersion of the tensor modes, and these results can be considered as an extension to plasmas and non-tensor gravitational modes in inhomogenenous backgrounds. In a sufficiently dense plasma, \textit{transverse} GEM modes consist of modes similar to the familiar transverse EM waves in plasma and the vacuum gravitational tensor modes. 
As compared to the vacuum GWs, the frequency shift of these waves due to ambient plasma is zero when the plasma is cold. When the plasma is hot, this shift is found to be generally of the same order as diffraction caused by plasma's contribution to the curvature of the background spacetime; therefore, it is beyond the accuracy of GO, or ray-optics \cite{book:tracy}, approximation, rendering the GO approximation invalid for studying such gravitational--EM wave interactions.\footnote{Some authors have reported small corrections to the dispersion relation of gravitational tensor modes by solving exact wave equations for specific backgrounds \cite{ref:chesters73,ref:madore73, ref:madore72, ref:barta18,ref:flauger18, ref:baym17, ref:asseo76}. These corrections are beyond GO, so the results are not local dispersion relations in the common sense and thus will not be considered below.} This result for a generic distribution of plasmas complements previous similar studies of neutral matter \cite{ref:chesters73,ref:madore73,ref:barta18,ref:flauger18, ref:baym17, ref:asseo76}, which have found negligible interaction of these tensor modes with neutral matter, and of homogeneous isotropic backgrounds which yield no interaction between the tensor modes and the vector EM modes. On the other hand, the \textit{longitudinal} GEM modes, which were studied in, for example, \citealp{ref:moretti19, ref:moretti20, ref:asenjo20}, can exhibit a strong interplay between gravitational and EM interactions in plasma. In the case of purely gravitational interactions, these longitudinal modes contain the so-called scalar modes that also yield the Newtonian Jeans mode in the limit of large refraction index \cite{my:gwjeans}. In particular, the dispersion relation of the Jeans mode is significantly affected by electrostatic interactions. As a spin-off, our calculation also provides an alternative resolution of the so-called Jeans swindle \cite{book:binney11, ref:falco13, tex:ershkovich11}, by approaching the Jeans instability rigorously from the standpoint of general relativity rather than Newtonian gravity.

Our paper is organized as follows. In \cref{sec:prelim}, we introduce the necessary basic concepts and notation. In \cref{sec:plasmaem}, we derive the oscillation-center Hamiltonian for a charged particle in a prescribed quasimonochromatic wave that involves metric perturbation and EM four-potential simultaneously. In \cref{sec:waveeqEM}, we use this result to derive gauge-invariant linear wave equations for self-consistent oscillations of gravitational and EM fields. We also comment on the gauge invariance of the dispersion operators, and ensure that the dispersion operator that we derive is consistent with the general theory in this regard. In \cref{sec:quasi}, we explore solutions of our equations in the limit of large refraction index, where GO applies. In particular, we discuss how electrostatic interactions affect the Jeans instability in plasma, and we also discuss the Jeans swindle. In \cref{sec:trans}, we study the interaction between the transverse  gravitational tensor modes and the transverse EM modes. In \cref{sec:conc}, we summarize the main results of our work.

\section{Preliminaries}
\label{sec:prelim}

\subsection{Notation}
\label{sec:notation}

Let us consider plasma in the presence of an EM field characterized by a four-potential \(\total{A}_\alpha\) and metric $\total{g}_{\alpha\beta}$ on a four-dimensional spacetime with coordinates \((x^0, x^1, x^2, x^3) \equiv x\) with signature \((-+++)\). Dynamics of this system is governed by the least-action principle
\begin{gather}\label{eq:S}
\delta S = 0, 
\qquad
S = S_{\rm m} + S_{\rm EM} + S_{\rm EH},
\end{gather}
where \(S_{\rm m}\) is the matter action,
\begin{gather}
S_{\rm EM} = -\frac{\varepsilon_0}{4}\int \dd^4x\,\sqrt{-\total{g}}\,\total{F}^{\alpha\beta}\total{F}_{\alpha\beta}
\end{gather}
is Maxwell's action, \m{\total{F}_{\alpha\beta} \doteq \nabla_\alpha \total{A}_\beta - \nabla_\beta \total{A}_\alpha} (the  symbol \m{\doteq} denotes definitions), \m{\total{g} \doteq \det \total{g}_{\alpha\beta}},
\begin{gather}
S_{\rm EH} = \frac{1}{2\kappa}\int \,\dd^4x\,\sqrt{-\total{g}}\,\total{R}
\end{gather}
is the Einstein--Hilbert action, \m{\total{R}} is the Ricci scalar, and $\kappa \doteq 8\pi G_{\rm N}/c^4$. We assume the same sign convention as in \citep{book:misner77}. We also assume geometrized Heaviside--Lorentz units, in which
\begin{gather}
c = 8\pi G_\text{N} = \varepsilon_0 = 1.
\end{gather}
Here, \m{c} is the speed of light, \m{G_\text{N}} is the Newtonian constant of gravitation, and \m{\varepsilon_0} is the vacuum permittivity.

We also assume that \(\total{A}_\alpha\) and $\total{g}_{\alpha\beta}$ can be decomposed (in the way to be specified shortly) as follows:
\begin{gather}
\total{A}_\alpha = A_\alpha + a_\alpha,
\qquad
\total{g}_{\alpha\beta} = g_{\alpha\beta} + h_{\alpha\beta}.
\label{eq:totalg}
\end{gather}
Here, $A_\alpha$ and $g_{\alpha\beta}$ are order-one background fields with a characteristic scale $\ell_g$, while $a_\alpha$ and $h_{\alpha\beta}$ are small perturbations:
\begin{gather}
a_\alpha = \mc{O}(\amag),
\qquad
h_{\alpha\beta} = \mc{O}(\hmag).
\end{gather}
The characteristic amplitude of the metric perturbation \m{\hmag \ll 1} is a dimensionless quantity. The characteristic amplitude of the EM four-potential \m{\amag} is a dimensional quantity that scales linearly with \m{\hmag} in self-consistent GEM waves. We also assume that the characteristic spacetime scale $\ell_h$ of these waves satisfies
\begin{subequations}
\begin{gather}
\epsilon \doteq \ell_h/\ell_g \ll 1.
\label{eq:golim}
\end{gather}
The existence of a small ``GO parameter'' \m{\epsilon} (whose availability should not always be taken for granted regardless of how small the wavelength is; see \cref{sec:trans}) allows us to introduce an intermediate scale \m{\ell_a} such that
\begin{gather}
\ell_h \ll \ell_a \ll \ell_g,
\label{eq:scalesep}
\end{gather}
\end{subequations}
and define the local average \(\favr{\ldots}\) over this scale.\footnote{Various averaging schemes can be used for this \cite{ref:brill64, ref:zalaletdinov96, tex:zalaletdinov97} and produce equivalent results \cite{ref:isi18, ref:caprini18, ref:riles13, ref:su12,ref:stein11} under the limit of scale separation \eqref{eq:scalesep}. For details about one possible implementation of the averaging, see \citep{my:gwquasi}, and a more general approach is presented in \citep{my:ql}.} Then, the splitting \eqref{eq:totalg} is specified by requiring that 
\begin{gather}\label{eq:havr}
\Braket{a_{\alpha}} = 0,
\qquad
\Braket{h_{\alpha\beta}} = 0,
\end{gather}
and, accordingly,
\begin{gather}\label{eq:gavr}
A_{\alpha} = \Braket{\total{A}_{\alpha}},
\qquad
g_{\alpha\beta} = \Braket{\total{g}_{\alpha\beta}}.
\end{gather}
In this paper, we will also assume that 
\begin{gather}
A_\alpha = 0.
\label{eq:Aavr}
\end{gather}
In particular, this means that the plasma is assumed to be non-magnetized. This assumption is adopted only to streamline the demonstration of our general approach to deriving GEM modes. Generalization to nonzero \m{\Braket{A_\alpha}} is conceptually straightforward and is left to future work.

\subsection{Approximate action}

Following the standard approach \cite{ref:cary81, my:itervar}, we also assume that the plasma responds adiabatically to gravitational and EM fields, meaning that all perturbations to plasma parameters can be unambiguously expressed through $h_{\alpha\beta}$ and $a_\alpha$ (and parameters of the unperturbed system). Then, using
\begin{gather}
h_{\alpha\beta} = \mc{O}(\hmag^1),
\qquad
a_{\alpha} = \mc{O}(\hmag^1),
\end{gather}
one can represent the total action $S$ [\cref{eq:S}] as a power series in \m{\hmag}:
\begin{gather}
S = \sum_n S^{(n)},
\qquad 
S^{(n)} = \mc{O}(\hmag^n).
\end{gather}
To the extent that terms of the third and higher orders in \m{\hmag} are negligible \cite{my:gwjeans}, this yields
\begin{gather}\label{eq:Savr0}
S \simeq S^{(0)}[g] + S^{(2)}[g, h, a],
\end{gather}
where the square brackets denote functional arguments (whose indices are omitted for brevity) and
\begin{gather}
S^{(2)} =  S^{(2)}_{\rm EM} + S^{(2)}_{\rm EH} + S^{(2)}_{\rm m}.
\label{eq:S2EM}
\end{gather}
Note that due to the scale separation \eqref{eq:scalesep}, the integrands in \m{S^{(n)}} can be replaced with their averaged values. For the same reason [and the fact that the integrand in \m{S^{(1)}} has zero average due to \cref{eq:havr}], the action \(S^{(1)}\) does not contribute to \cref{eq:Savr0}. Within linear theory, this treatment already yields a gauge-invariant system, as shown in our \citep{my:gwquasi}, where a more general approach to maintaining gauge invariance is presented using the gauge-invariant variables introduced in our \citep{my:gwinvar}.

As seen easily, the second-order EM action is given by
\begin{gather}
 S^{(2)}_{\rm EM} = - \frac{1}{4} \int \dd^4x\,\sqrt{-g}\,
 (\nabla^\alpha a^{\beta} - \nabla^\beta a^{\alpha})
 (\nabla_\alpha a_{\beta} - \nabla_\beta a_{\alpha}),
\label{eq:LEM}
\end{gather}
where $g \doteq \det g_{\alpha\beta}$ and $\nabla$ is the covariant derivative associated with the background metric $g_{\alpha\beta}$. (From now on, the indices of $a_\alpha$ and $h_{\alpha\beta}$ are also manipulated using the background metric, as usual.) Also, as shown in \citep{my:gwjeans} and references cited therein, one has
\begin{gather}
S_{\rm EH}^{(2)}  = \int \dd^4 x\,\left(\mc{L}_G^{(2)} + \mc{L}_{\rm vac}^{(2)}\right),
\\
S^{(2)}_{\rm m} = \int \dd^4x\,\mc{L}_{\rm m}^{(2)},
\end{gather}
where
\begin{align}
&\mc{L}_G^{(2)}
\doteq \frac{\sqrt{-g}}{4}\bigg(
- \frac{1}{2}\, R h^{\alpha\beta} h_{\alpha\beta}
- R_{\alpha\beta} h^{\alpha\beta} h
\notag \\&\qquad\qquad\qquad\qquad
+ \frac{1}{4}\, R h^2
+ 2 R_{\alpha\beta} h^{\alpha\rho} {h_\rho}^\beta
\bigg),
\label{eq:LG}
\\
&\mc{L}^{(2)}_{\rm vac} \doteq \frac{\sqrt{-g}}{4}
\bigg(
- \frac{1}{2}\, \nabla^\rho h^{\alpha\beta} \nabla_\rho h_{\alpha\beta}
+ \frac{1}{2}\, \nabla^\rho h \nabla_\rho h
\notag \\&\qquad\qquad\qquad
- \nabla_\alpha h \nabla_\beta h^{\alpha\beta}
+ \nabla^\rho h^{\alpha\beta} \nabla_\alpha h_{\beta\rho}
\bigg),
\label{eq:Lvac}
\end{align}
$R_{\alpha\beta}$ is the Ricci tensor associated with the background metric, and \m{\mc{L}_{\rm m}^{(2)}} is the second-order Lagrangian density of the matter. Under the assumption of scale separation \eqref{eq:scalesep}, one can as well replace \m{\mc{L}_{\rm m}^{(2)}} with its spatial average, \m{\favr{\mc{L}_{\rm m}^{(2)}}}. Below, we discuss how to calculate the latter in detail.

\section{Matter action}
\label{sec:plasmaem}

\subsection{Fluid model}

As a preliminary step to considering actual plasma, let us calculate the average Lagrangian density of a single relativistic fluid in prescribed fields, assuming that the fluid consists of particles with masses~$m$ and charges~$e$. As usual, spin effects are considered negligible. Then, following \citep{my:gwponder} (see also \citealp{my:qponder, my:lens}), the corresponding Lagrangian density can be obtained as the semiclassical limit of the Klein--Gordon Lagrangian density
\begin{multline}\label{eq:Lm}
\mc{L}_{\rm m} = \frac{\sqrt{-\total{g}}}{2m}
\Big[
-\total{g}^{\alpha\beta}
\left(\pd_{\alpha} \psi^* + \ii e \total{A}_\alpha \psi^*\right)
\left(\pd_{\beta} \psi - \ii e \total{A}_\beta \psi \right)
\\
- m^2|\psi|^2
\Big],
\end{multline}
where \m{\psi} is a quantum mean field normalized such that \m{\mc{I} \doteq |\psi|^2} is the local number density. (For the correspondence between quantum and classical variational principles, see \citep{my:qlagr}.) Let us express this wavefunction in the Madelung form, $\psi = \sqrt{\mc{I}}\, \ee^{\ii\vartheta}$, where $\vartheta$ is a real phase. In the semiclassical limit, in which 
\begin{gather}
p_\alpha \doteq \nabla_\alpha \vartheta \gg \nabla_\alpha \ln{\mc{I}},
\end{gather}
\Cref{eq:Lm} can be simplified as follows:
\begin{gather}
\mc{L}_{\rm m} = -\sqrt{-\total{g}}\, \mc{I}(x) H(x, \nabla\vartheta).
\end{gather}
Here, the function \m{H(x, p)} is given by
\begin{gather}\label{eq:HEM}
H(x, p)
= \frac{1}{2m}\,[\total{g}^{\alpha\beta}
(p_\alpha - e \total{A}_\alpha) (p_\beta - e \total{A}_\beta) + m^2]
\end{gather}
and can be recognized as a Hamiltonian of the particle dynamics in spacetime. 

Let us expand $H$ in powers of \m{\hmag} and, as before, keep terms only up to $\mc{O}(\hmag^2)$. This gives \(H \simeq H^{(0)} + H^{(1)} + H^{(2)}\), where $H^{(n)} = \mc{O}(\hmag^n)$. Using \cref{eq:Aavr} and \(\total{g}^{\alpha\beta} = g^{\alpha\beta} - h^{\alpha\beta} + {h^\alpha}_\gamma h^{\gamma\beta} + \mc{O}(\hmag^3)\), one readily finds that
\begin{subequations}
\begin{gather}
H^{(0)} = \frac{1}{2m} \left(g^{\alpha\beta} p_\alpha p_\beta + m^2 \right),
\label{eq:H0EM}
\\
H^{(1)} = -\frac{1}{2m}\,h^{\alpha\beta}p_\alpha p_\beta
- \frac{e}{m}\, p^\alpha a_\alpha,
\\
H^{(2)}=\frac{1}{2m}\,{h^\alpha}_\gamma h^{\gamma\beta}p_\alpha p_\beta
+ \frac{e}{m}\, h^{\alpha\beta} p_\alpha a_\beta
+ \frac{e^2}{2m}\, g^{\alpha\beta} a_\alpha a_\beta.
\end{gather}
\label{eq:H12EM}%
\end{subequations}
Next, let us assume that the EM field and the metric perturbation are quasimonochromatic, \ie
\begin{gather}
h_{\alpha\beta} = \Re (\ee^{\ii \theta} \hamp_{\alpha\beta}),
\quad
a_\alpha = \Re (\ee^{\ii \theta} \aamp_\alpha),
\label{eq:mono}
\end{gather}
where \(\theta\) is a rapid phase and \(\hamp_{\alpha\beta}\) and \(\aamp_\alpha\) are slow envelopes, with the local wavevector defined by
\begin{gather}
k_\alpha \doteq \pd_\alpha \theta = \nabla_\alpha \theta \sim \ell_h^{-1}.
\end{gather}
Then, like in \citep{my:gwponder}, it can be shown that 
\begin{gather}
\favr{\mc{L}_{\rm m}} = -\sqrt{-g}\,\bar{\mc{I}}  \mc{H}(x, \nabla\bar{\vartheta}),
\label{eq:avglagft}
\end{gather}
where \(\sqrt{-g}\, \bar{\mc{I}} \doteq \Braket{\sqrt{-\total{g}}\,\mc{I}}\), \m{\bar{\vartheta} \doteq \favr{\vartheta}}, and \(\mc{H}=H^{(0)}+\mc{H}^{(2)}\), with
\begin{gather}
\mc{H}^{(2)} = \favr{H^{(2)}} - \frac{m k_\mu}{2}\frac{\pd}{\pd p_\mu}
\bigg(
\frac{\favr{{(H^{(1)}})^2}}{k_\lambda p^\lambda}
\bigg).
\label{eq:ponderhamiltonian0}
\end{gather}
(Here, we assume nonresonant particles, \ie particles with \m{k_\lambda p^\lambda} sufficiently far from zero, so that \m{\mc{H}^{(2)} \ll H^{(0)}}. Resonant particles can be rigorously accommodated within more comprehensive approaches, such as those in \citep{my:ql, my:nonloc} or simply as in homogeneous-plasma wave theory \cite{book:stix}. The corresponding modifications of the theory are obvious; see \cref{sec:df}.)

One can understand \m{\mc{H}} as the effective, or ``oscillation-center'' (OC), Hamiltonian that governs the average dynamics of particles in spacetime \cite{ref:dewar73, my:qponder, my:ql, my:gwponder}. (The term OC denotes a fictitious particle whose trajectory coincides with the average trajectory of the actual particle.) The corresponding Euler--Lagrange equations can be found by considering the variation of the action with respect to \(\bar{\mc{I}}\) and \(\bar{\vartheta}\), yielding
\begin{subequations}
\begin{gather}
\mc{H}(x, \nabla\bar{\vartheta})=0,
\label{eq:euler1ponder}
\\
\frac{\pd}{\pd x^\alpha}\left[\sqrt{-g} \,
\bar{\mc{I}}(x) \,
\frac{\pd\mc{H}(x, \bar{\vartheta})}{\pd p_\alpha}
\right] = 0,
\label{eq:euler2ponder}
\end{gather}%
\end{subequations}
where \cref{eq:euler1ponder} determines the energy--momentum relation (on-shell condition) for an OC, and \cref{eq:euler2ponder} represents the continuity equation for OCs. 

A direct calculation shows that
\begin{align}
\mc{H}^{(2)}(x, p) = &
\,\frac{\Braket{h_{\alpha\beta} h_{\gamma\delta}}}{2m}\, g^{\beta\gamma} p^\alpha p^\delta
\notag\\
& + \frac{e}{m} \Braket{a_\alpha h_{\beta\gamma}} g^{\alpha\gamma} p^\beta
\notag\\
& + \frac{e^2}{2m} \Braket{a_\alpha a_\beta} g^{\alpha\beta}
\notag\\
& - \Braket{h_{\alpha\beta} h_{\gamma\delta}} \frac{k_\mu}{8m}\frac{\pd}{\pd p_\mu}
\bigg(
\frac{p^\alpha p^\beta p^\gamma p^\delta}{k_\lambda p^\lambda}
\bigg)
\notag\\
& - e \Braket{a_\alpha h_{\beta\gamma}} \frac{k_\mu}{2m}\frac{\pd}{\pd p_\mu}
\bigg(
\frac{p^\alpha p^\beta p^\gamma}{k_\lambda p^\lambda}
\bigg)
\notag\\
& - e^2 \Braket{a_\alpha a_\beta} \frac{k_\mu}{2m}\frac{\pd}{\pd p_\mu}
\bigg(
\frac{p^\alpha p^\beta}{k_\lambda p^\lambda}
\bigg).
\label{eq:H2}
\end{align}
Then, the part of \m{\favr{\mc{L}_{\rm m}}} that is of the second order in the wave amplitude can be expressed as follows:
\begin{gather}
\favr{\mc{L}^{(2)}_{\rm m}} = -\sqrt{-g} N \Phi,
\label{eq:Lf2}
\end{gather}
where $N \doteq \bar{\mc{I}} p^0/m$ and \m{\Phi \doteq m \mc{H}^{(2)}/p^0}. The function $N$ can be understood as the OC number density\footnote{To the leading order, \cref{eq:euler2ponder} can be viewed as the continuity equation for \(N\), where we ignored \(\mc{O}(\hmag^2)\) terms and replaced \(\pd \mc{H}/\pd p_\alpha\) with \(p^\alpha/m\) \eqref{eq:H0EM}.} but, within linear-wave theory, does not need to be distinguished from the unperturbed number density of the fluid. The function \m{\Phi} is understood as the second-order part of OC's \textit{spatial}-dynamics Hamiltonian (as opposed to the spacetime-dynamics Hamiltonian), or, loosely, the ``ponderomotive potential'' \cite{my:gwponder}. Using \cref{eq:H2}, it can also be written~as 
\begin{align}
\Phi 
= & \; \frac{1}{2p^0}\,\favr{h_{\alpha\beta} h_{\gamma\delta}} \;
\left[
g^{\beta\gamma} p^\alpha p^\delta
-\frac{k_\mu}{4}\frac{\pd}{\pd p_\mu}
\left(
\frac{p^\alpha p^\beta p^\gamma p^\delta}{k_\lambda p^\lambda}
\right)
\right]
\notag \\
& + \frac{e}{p^0} \,\favr{a_\alpha h_{\beta\gamma}} \,
\left[
g^{\alpha\gamma} p^\beta
-\frac{k_\mu}{2}\frac{\pd}{\pd p_\mu}
\left(
\frac{p^\alpha p^\beta p^\gamma}{k_\lambda p^\lambda}
\right)
\right]
\notag \\
& + \frac{e^2}{2p^0}\,\favr{a_\alpha h_{\beta\gamma}}
\left[
g^{\alpha\beta}
- k_\mu\frac{\pd}{\pd p_\mu}
\bigg(
\frac{p^\alpha p^\beta}{k_\lambda p^\lambda}
\bigg)
\right],
\label{eq:Phi}
\end{align}
a form that will be more convenient below.

\subsection{General case}

The general case, when plasma consists of multiple species with general distributions of momenta, can be considered as the case of multiple fluids that contribute to \m{\favr{\mc{L}^{(2)}_{\rm m}}} additively. Suppose particles of type $s$, with masses $m_s$ and charges $e_s$, are characterized by a distribution function~\m{f_s}. (A single particle corresponds to a delta-shaped $f_s$.) As usual in plasma theory, we assume that this function is normalized such that
\begin{gather}
\int \dd \boldsymbol{p}\,f_s(x, \boldsymbol{p}) = N_s(x)
\end{gather}
(here, \m{x \equiv (t, \vec{x})} is the four-dimensional spacetime coordinate, $t$ is time, $\vec{x}$ is the three-dimensional spatial coordinate, and \m{\boldsymbol{p}} is the three-dimensional spatial momentum), or equivalently,
\begin{gather}
\int \frac{\dd\boldsymbol{p}}{p^0}\,f_s(x, \boldsymbol{p}) = \frac{\rho_s(x)}{m_s^2},
\end{gather}
where $\rho_s(x)$ is the local proper mass density. Then, 
\begin{align}
\favr{\mc{L}_{\rm m}^{(2)}} 
& = - \sqrt{-g} \sum_s N_s \Phi_s 
\notag\\
& = - \sqrt{-g} \sum_s \int \dd \boldsymbol{p}\,\Phi_s f_s(x, \boldsymbol{p}).
\end{align}
Then, using \cref{eq:Phi}, one obtains
\begin{align}
S_{\rm m}^{(2)} = & - \frac{1}{2} \sum_s \int \dd^4x\,\sqrt{-g} \int \frac{\dd \boldsymbol{p}}{p^0} \, f_s(x,\boldsymbol{p}) 
\notag\\
& \times 
\bigg\{
h_{\alpha\beta} h_{\gamma\delta} \left[
g^{\beta\gamma} p^\alpha p^\delta
-\frac{k_\mu}{4}\frac{\pd}{\pd p_\mu}
\left(
\frac{p^\alpha p^\beta p^\gamma p^\delta}{k_\lambda p^\lambda}
\right)
\right]
\notag\\
& \hphantom{\times \bigg\lbrace} + 2 e_s a_\alpha h_{\beta\gamma}
\left[
g^{\alpha\gamma} p^\beta
-\frac{k_\mu}{2}\frac{\pd}{\pd p_\mu}
\left(
\frac{p^\alpha p^\beta p^\gamma}{k_\lambda p^\lambda}
\right)
\right]
\notag\\
& \hphantom{\times \bigg\lbrace} + e_s^2 a_\alpha a_\beta
\left[
g^{\alpha\beta}
- k_\mu\frac{\pd}{\pd p_\mu}
\bigg(
\frac{p^\alpha p^\beta}{k_\lambda p^\lambda}
\bigg)
\right]
\bigg\},
\label{eq:L2m}
\end{align}
where the space averaging on the right-hand side is dropped as it has no impact on the integral within the GO limit. Also note that the first term in the third line above ($\propto g^{\alpha\gamma} p^\beta$) provides zero contribution to \m{S_{\rm m}^{(2)}}, because the assumed absence of background EM fields [\cref{eq:Aavr}] implies neutrality of the background plasma and the absence of background currents, \ie
\begin{gather}
\sum_s \int \frac{\dd \boldsymbol{p}}{p^0} \, f_s(x,\boldsymbol{p}) \, e_s \, p^\alpha = 0.
\label{eq:backj}
\end{gather}

\subsection{Dispersion functions}
\label{sec:df}

Let us introduce the following inner product for any pair of fields $u_1$ and $u_2$ on the background space:
\begin{gather}\label{eq:ip}
\Braket{u_1, u_2} = \int \dd^4 x\,\sqrt{-g}\,u_1(x) u_2(x).
\end{gather}
Using this notation, the matter action \eqref{eq:L2m} can be expressed as follows:
\begin{multline}
S^{(2)}_{\rm m}
= \frac{1}{2} \Big(
\Braket{h^{\alpha\beta}, D_{\alpha\beta\gamma\delta}^{\rm mG} h^{\gamma\delta}}
\\
+ 2 \Braket{a^{\alpha}, D_{\alpha\beta\gamma}^{\rm mGEM} h^{\beta\gamma}}
+ \Braket{a^{\alpha}, D_{\alpha\beta}^{\rm mEM} a^{\beta}}
\Big),
\label{eq:S2m}
\end{multline}
and
\begin{subequations}
\begin{align}
D_{\alpha\beta\gamma\delta}^{\rm mG}
& \doteq \sum_s \int \frac{\dd \boldsymbol{p}}{p^0} \, f_s(x,\boldsymbol{p})
\bigg[
\frac{k_\mu}{4}\frac{\pd}{\pd p_\mu}
\left(
\frac{\mc{T}_{\alpha\beta} \mc{T}_{\gamma\delta}}{\Omega}
\right)
\nonumber \\ & \qquad\qquad\qquad\qquad\qquad\qquad
- Q_{\alpha\beta\gamma\delta}
\bigg],
\label{eq:DmG}
\\[1ex]
D_{\alpha\beta\gamma}^{\rm mGEM}
& \doteq \sum_s \int \frac{\dd \boldsymbol{p}}{p^0} \, f_s(x,\boldsymbol{p}) e_s
\bigg[\frac{k_\mu}{2}\frac{\pd}{\pd p_\mu}
\left(
\frac{\mc{T}_{\alpha\beta} p_\gamma}{\Omega}
\right)
\nonumber \\ & \qquad\qquad\qquad\qquad\qquad\quad\:\,
- g_{\alpha(\gamma} p_{\beta)}
\bigg],
\label{eq:DmGEM}
\\[1ex]
D_{\alpha\beta}^{\rm mEM}
& \doteq \sum_s \int \frac{\dd \boldsymbol{p}}{p^0} \, f_s(x,\boldsymbol{p}) e^2_s
\bigg[
k_\mu\frac{\pd}{\pd p_\mu}
\bigg(
\frac{\mc{T}_{\alpha\beta}}{\Omega}
\bigg)
\nonumber \\ & \qquad\qquad\qquad\qquad\qquad\qquad\quad
- g_{\alpha\beta}
\bigg],
\label{eq:DmEM}
\end{align}
\label{eq:DGEM}%
\end{subequations}
where we introduced
\begin{gather}
Q_{\alpha\beta\gamma\delta} \doteq \frac{1}{4} \left(g_{\beta\gamma}  \mc{T}_{\alpha\delta} + g_{\alpha\delta} \mc{T}_{\beta\gamma} + g_{\alpha\gamma} \mc{T}_{\beta\delta} + g_{\beta\delta}  \mc{T}_{\alpha\gamma}
\right),
\\
\Omega_{\alpha\beta} \doteq p_{(\alpha} k_{\beta )},
\quad
\Omega \doteq p_\alpha k^\alpha,
\\
\mc{T}_{\alpha\beta} \doteq p_\alpha p_\beta.
\end{gather}
(Note that \(\mc{X}_{\alpha\beta\gamma\delta}\) introduced in \cite[Eq.~(120)]{my:gwponder} is the same as \(D^{\rm mG}_{\alpha\beta\gamma\delta}\) in \cref{eq:DmG}, as can be seen by comparing \cref{eq:DmG} with \cite[Eq.~(B1)]{my:gwponder}.) The functions \eqref{eq:DGEM} will be called dispersion functions.\footnote{In a more general sense, these are the Weyl symbols of the corresponding dispersion operators \cite{my:quasiop1}. However, in the GO limit considered here, these subtleties can be ignored.} In particular, notice that the last term in the square bracket in \cref{eq:DmGEM} vanishes because of \cref{eq:backj}; however, it is retained here to accentuate parallels between the three expressions \eqref{eq:DGEM}. Finally, following the same steps as in \cite[Appendix~B]{my:gwponder}, one also obtains the following alternative representations of the dispersion functions to be used below
\begin{subequations}
\begin{align}
D_{\alpha\beta\gamma\delta}^{\rm mG}
& \doteq \sum_s \int  \frac{\dd\boldsymbol{p}}{4(p^0)^2}
\bigg(
\frac{\boldsymbol{k}\cdot \pd_{\boldsymbol{p}} f_s}{\omega - \boldsymbol{k} \cdot \boldsymbol{v}}\,
\mc{T}_{\alpha\beta} \mc{T}_{\gamma\delta}
\nonumber \\ & \qquad\qquad\qquad\qquad\qquad\quad
+ f_s J^{\rm mG}_{\alpha\beta\gamma\delta} 
\bigg),
\\[1ex]
D_{\alpha\beta\gamma}^{\rm mGEM}
& \doteq \sum_s \int  \frac{\dd\boldsymbol{p}}{2(p^0)^2}
\, e_s
\bigg(
\frac{\boldsymbol{k}\cdot \pd_{\boldsymbol{p}} f_s}{\omega - \boldsymbol{k} \cdot \boldsymbol{v}}\,
\mc{T}_{\alpha\beta} p_\gamma
\nonumber \\ & \qquad\qquad\qquad\qquad\qquad\:
+ f_s J^{\rm mGEM}_{\alpha\beta\gamma} 
\bigg),
\\[1ex]
D_{\alpha\beta}^{\rm mEM}
& \doteq \sum_s \int  \frac{\dd\boldsymbol{p}}{(p^0)^2}
\, e^2_s
\bigg(
\frac{\boldsymbol{k}\cdot \pd_{\boldsymbol{p}} f_s}{\omega - \boldsymbol{k} \cdot \boldsymbol{v}}\,
\mc{T}_{\alpha\beta}
\nonumber \\ & \qquad\qquad\qquad\qquad\qquad\quad
+ f_s J^{\rm mEM}_{\alpha\beta} 
\bigg),
\end{align}
\label{eq:DGEMreso}
\end{subequations}
where
\begin{subequations}
\begin{align}
J^{\rm mG}_{\alpha\beta\gamma\delta} 
\doteq &\;
\frac{\pd(\mc{T}_{\alpha\beta} \mc{T}_{\gamma\delta})}{\pd p_0}
-
\frac{g^{00}}{p^0}\,\mc{T}_{\alpha\beta} \mc{T}_{\gamma\delta}
- 4 p^0 Q_{\alpha\beta\gamma\delta},
\\
J^{\rm mGEM}_{\alpha\beta\gamma} 
\doteq &\;
\frac{\pd(\mc{T}_{\alpha\beta} p_\gamma)}{\pd p_0}
-
\frac{g^{00}}{p^0}\,\mc{T}_{\alpha\beta} p_\gamma
- 2 p^0 g_{\alpha(\gamma} p_{\beta)},
\\
J^{\rm mEM}_{\alpha\beta} 
\doteq &\;
\frac{\pd\mc{T}_{\alpha\beta}}{\pd p_0}
-
\frac{g^{00}}{p^0}\,\mc{T}_{\alpha\beta}
- p^0 g_{\alpha\beta}.
\end{align}
\label{eq:df2}%
\end{subequations}

As a reminder, the above calculation assumes that no plasma particles are resonant to the wave; \ie \(f_s (x, \boldsymbol{p}) = 0\) at those \m{\boldsymbol{p}} that satisfy \(\Omega = 0\) at given $x$. However, resonant particles can be added into consideration in the same way as it is commonly done in theory of plasma dispersion \cite{book:stix, my:nonloc}. (For the most comprehensive treatment, see \citep{my:ql}.) Then, \autoref{eq:df2} remain valid for \(\Im\omega > 0\), and analytic continuation of the corresponding expressions for the dispersion functions should be used otherwise. These analytic continuations can be obtained by integrating over the Landau contour as opposed to the real momentum space \cite{book:stix}.

Also, as a side remark, note that within the model considered here, EM and gravitational perturbations couple only via plasma and vanish in the limit when the plasma density is negligible. This is because the direct contribution of the photons to the spacetime curvature is assumed negligible compared to that of massive plasma particles. This means that the coupling considered here is different from the known direct photon--graviton conversion in the presence of magnetic fields \cite{ref:fujita20, ref:gertsenshtein62}.

\section{Equation for gravito-electromagnetic waves}
\label{sec:waveeqEM}

\subsection{Assumptions}
\label{sec:asm}

In what follows, we use normal coordinates, in which the first-order derivatives of the background metric vanish. Then,
\begin{gather}
g_{\alpha\beta} = \eta_{\alpha\beta} + \frac{1}{2}\, (\pd_\sigma\pd_\rho g_{\alpha\beta})\, x^\rho x^\sigma + \mc{O}(\ell_g^{-3}),
\end{gather}
and metric's second-order derivatives can be expressed through the Riemann tensor \m{R_{\alpha\beta\gamma\delta}} as \cite{ref:brewin98}
\begin{gather}
\pd_\sigma \pd_\rho g_{\alpha\beta}
= - \frac{1}{3} \left(
R_{\alpha\rho\beta\sigma} + R_{\alpha\sigma\beta\rho}
\right).
\end{gather}
As discussed in \citep{my:gwjeans}, the interaction of GWs with matter (in our case, plasma) is significant only when the Weyl tensor is not the dominant contributor to the curvature and \(R_{\alpha\beta}\), which is \(\mc{O}(\rho)\) by Einstein field equations, is of the same order as \(R_{\alpha\beta\gamma\delta}\), which is \(\ell_g^{-2}\). Thus, below we assume that
\begin{gather}
\hmag \ll \epsilon \sim \ell_h/\ell_g \sim \ell_h \sqrt{\rho} \ll 1,
\label{eq:scaleEM}
\end{gather}
where $\rho$ is the total mass density. Note that, in typical plasmas, the ions are the dominant contributors to the mass density. That is,
\begin{gather}
\rho \simeq \rho_i \ll \rho_e,
\end{gather}
where the index \(i\) denotes ions, the index \(e\) denotes electrons, and \(m_i \gg m_e\). (For simplicity, we assume that, in addition to electrons, plasma contains ions of only one type.) Our results are also applicable at smaller densities, but then wave's gravitational coupling with matter is beyond the accuracy of our approximation and thus will be neglected.

Another small parameter that we will use is \(m_s/e_s\). In typical plasmas, the values of \m{m_s/e_s} for various species are very small compared to unity. For example, electrons have \(m_e/e_e \sim 10^{-21}\), and protons (denoted with index~\(p\)) have \(m_p/e_p \sim 10^{-18}\). We will also assume
\begin{gather}
\frac{m_s^2}{e_s^2} \ll \epsilon
\label{eq:scaleEM1}
\end{gather}
for all species. Like in the case with \cref{eq:scaleEM}, our results are also applicable at arbitrarily small \m{\epsilon}, which corresponds to vanishingly small densities, but then wave's gravitational coupling with matter must be neglected.

\subsection{Euler--Lagrange equations for the EM field}
\label{sec:eulerem}

The first set of GEM equations is obtained by considering the variation of the action \eqref{eq:S2EM} with respect to the EM four-potential:
\begin{gather}
0 =\frac{\delta S^{(2)}}{\delta a_\alpha}
= \frac{\delta S^{(2)}_{\rm EM}}{\delta a_\alpha}
+ \frac{\delta S^{(2)}_{\rm m}}{\delta a_\alpha}.
\label{eq:waveA0}
\end{gather}
The first term on the right-hand side is well known \cite[Sec.~30]{book:landau2}:
\begin{gather}
\frac{g_{\alpha\beta}}{\sqrt{-g}}\frac{\delta S^{(2)}_{\rm EM}}{\delta a_\beta}
= - {a_{\beta,\alpha}}^{\beta} + {a_{\alpha,\beta}}^\beta.
\label{eq:maxvar}
\end{gather}
Using \cref{eq:mono}, this can also be expressed as
\begin{multline}
\frac{g_{\alpha\beta}}{\sqrt{-g}}\frac{\delta S^{(2)}_{\rm EM}}{\delta a_\beta}
= k_\alpha k^\beta a_\beta
- g^{\beta\gamma}\mcc{R}^{\rm EM}_{\beta\alpha\gamma}
\\
- k^2 a_\alpha + g^{\beta\gamma} \mcc{R}^{\rm EM}_{\alpha\beta\gamma},
\end{multline}
where we have introduced
\begin{gather}
\mcc{R}^{\rm EM}_{\alpha\beta\gamma} \doteq \ee^{\ii \theta} (\aamp_{\alpha,\beta\gamma} + \ii k_\gamma \aamp_{\alpha,\beta} + \ii k_\beta \aamp_{\alpha,\gamma} + \ii k_{\gamma,\beta} \aamp_\alpha).
\end{gather}
Note that \m{\mcc{R}^{\rm EM}_{\alpha\beta\gamma} \sim \epsilon}, while 
\begin{gather}
D^{\rm mEM}_{\alpha\beta} \sim e^2\rho_e/m_e^2 \sim e^2 \epsilon^2/m_e m_i,
\end{gather}
where $e = |e_e| = -e_e$ is the elementary charge, and we used \(\rho_e/\rho_i \sim m_e/m_i\) due to the background-plasma neutrality. Then, under the assumption \eqref{eq:scaleEM1}, \m{\mcc{R}^{\rm EM}_{\alpha\beta\gamma}} can be ignored compared with \m{D^{\rm mEM}_{\alpha\beta}}. This leads to
\begin{gather}
\frac{g_{\alpha\beta}}{\sqrt{-g}}\frac{\delta S^{(2)}_{\rm EM}}{\delta a_\beta}
\simeq k_\alpha k^\beta a_\beta
- k^2 a_\alpha.
\label{eq:aux1}
\end{gather}
In conjunction with \autoref{eq:DGEM} or \eqref{eq:DGEMreso} for \(S^{(2)}_{\rm m}\), \cref{eq:aux1} can be used to rewrite \cref{eq:waveA0} as
\begin{gather}
k_\alpha k^\beta a_\beta
- k^2 a_\alpha
+ D^{\rm mEM}_{\alpha\beta} a^\beta
+ D^{\rm mGEM}_{\alpha\beta\gamma} h^{\beta\gamma}
= 0.
\label{eq:waveA1}
\end{gather}
\Cref{eq:waveA1} determines the GO dispersion relation and the local polarization of GEM waves. (For this, it must be combined with the equation for \m{h^{\beta\gamma}}, which is derived in \cref{sec:waveeqh}.) In the absence of gravitational interactions, \cref{eq:waveA1} coincides with the well-known equation for dispersive waves in relativistic nonmagnetized plasma, which is usually expressed through the corresponding dielectric tensor; see, \eg Eq.~(9.42) in \citep{my:ql}. 

Note that, although approximate, \cref{eq:waveA1} honors symmetries of the exact equations that describe the whole system. Indeed, as one can check by direct calculation [with \cref{eq:backj} taken into account], the dispersion functions satisfy
\begin{subequations}
\begin{gather}
k^\alpha D^{\rm mEM}_{\alpha\beta} = 0,
\label{eq:kDmEM}
\\
k^\alpha D^{\rm mGEM}_{\alpha\beta\gamma} = 0,
\label{eq:kaDmGEM}
\\
k^\beta D^{\rm mGEM}_{\alpha\beta\gamma} = 0.
\label{eq:kbDmGEM}
\end{gather}
\label{eq:kDGEM}
\end{subequations}
From here, one can see that \cref{eq:waveA1} is invariant with respect to the EM gauge transformations
\begin{gather}
a_\alpha \to a_\alpha + k_\alpha \varphi,
\label{eq:gaugeEM}
\end{gather}
where \(\varphi\) is any scalar function. \Cref{eq:kbDmGEM} also ensures invariance of \cref{eq:waveA1} with respect to the gauge transformation of the metric perturbation, 
\begin{gather}
h_{\alpha\beta} \to h'_{\alpha\beta} = h_{\alpha\beta} - \lie_\xi g_{\alpha\beta},
\label{eq:gaugeG}
\end{gather}
where $\lie_\xi g_{\alpha\beta} = 2 \Lambda_{(\alpha} k_{\beta)} \ee^{\ii\theta}$ is the Lie derivative of the background metric with respect to an arbitrary vector wave field \(\xi^\alpha = \Re(-\ii \Lambda^\alpha \ee^{\ii\theta})  = \mc{O}(\hmag)\) in the GO limit.

Equations~\eqref{eq:kDGEM} can be considered as a test for the accuracy of approximate models of the GW dispersion in plasma, like in \cite[Eq.~(4.20)]{my:gwjeans}. As shown in \citep{my:gwjeans}, 
certain terms in the dispersion operator may not pass this test if derived under the GO approximation and, thus, contain spurious coordinate effects that should be dealt with.

\subsection{Euler--Lagrange equations for the metric perturbation}
\label{sec:waveeqh}

The other set of GEM equations can be obtained by considering the variation of the action \eqref{eq:S2EM} with respect to the metric perturbation:
\begin{gather}
\frac{\delta S^{(2)}}{\delta h^{\alpha\beta}} = 0.
\end{gather}
These equations can be expressed as
\begin{gather}
\left(
\widehat{D}^{\rm vac}_{\alpha\beta\gamma\delta}
+ \widehat{\mc{G}}_{\alpha\beta\gamma\delta}
+ D^{\rm mG}_{\alpha\beta\gamma\delta}
\right) h^{\gamma\delta}
+ a^\mu D^{\rm mGEM}_{\mu\alpha\beta}= 0,
\label{eq:aux5}
\end{gather}
with the operators \(\widehat{D}^{\rm vac}_{\alpha\beta\gamma\delta}\), and \(\widehat{\mc{G}}_{\alpha\beta\gamma\delta}\) defined as \cite{my:gwjeans}
\begin{multline}
\widehat{D}^{\rm vac}_{\alpha\beta\gamma\delta} h^{\gamma\delta} \doteq
\frac{1}{4}
\big(
\pd^\rho \pd_\rho h_{\alpha\beta}
- g_{\alpha\beta} g^{\rho\sigma}\pd^\lambda \pd_\lambda h_{\rho\sigma}
+ g^{\rho\sigma} \pd_\alpha \pd_\beta h_{\rho\sigma}
\\+ g_{\alpha\beta} \pd^\rho \pd^\sigma h_{\rho\sigma}
- \pd^\rho \pd_\alpha h_{\beta\rho}
- \pd^\rho \pd_\beta h_{\alpha\rho} \big),
\end{multline}
\begin{multline}
\widehat{\mc{G}}_{\alpha\beta\gamma\delta}h^{\gamma\delta} \doteq
\frac{1}{4}\big(
- G h_{\alpha\beta}
- G_{\alpha\beta} h
+ 2 G_{\alpha\rho} {h_\beta}^\rho
\\+ 2 G_{\rho\beta} {h^\rho}_\alpha
- 2 g_{\alpha\beta} R_{\rho\sigma} h^{\rho\sigma}
+ 2 R_{\rho\alpha\sigma\beta} h^{\rho\sigma}
\big).
\label{eq:mcG}
\end{multline}
For quasimonochromatic waves [\cref{eq:mono}], \cref{eq:aux5} can also be written as
\begin{gather}
4 D^{(0)}_{\alpha\beta\gamma\delta} h^{\gamma\delta} + M_{\alpha\beta} + 4 a^\mu D^{\rm mGEM}_{\mu\alpha\beta} = 0.
\label{eq:euler1EM}
\end{gather}
Here, \(D^{(0)}_{\alpha\beta\gamma\delta}\) is defined as
\begin{multline}
D^{(0)}_{\alpha\beta\gamma\delta} \doteq \frac{1}{4} \,\big(
- k^2 g_{\alpha\gamma}g_{\beta\delta}
+ g_{\alpha\beta}g_{\gamma\delta} k^2
- k_\alpha k_\beta g_{\gamma\delta}
\\- g_{\alpha\beta} k_\gamma k_\delta
+ k_\alpha k_\gamma g_{\beta\delta}
+ k_\beta  k_\delta g_{\alpha\gamma}
\big),
\end{multline}
\(M_{\alpha\beta}\) describes wave's gravitational coupling to the matter,
\begin{align}
M_{\alpha\beta}
\doteq
& \,\,{\mcc{R}_{\alpha\beta\rho}}^\rho
- g_{\alpha\beta} g^{\rho\sigma} {\mcc{R}_{\rho\sigma\lambda}}^\lambda
+ g^{\rho\sigma} \mcc{R}_{\rho\sigma\alpha\beta}
\notag\\
& + g_{\alpha\beta} {\mcc{R}_{\rho\sigma}}^{\rho\sigma}
- {\mcc{R}_{\rho\beta\alpha}}^\rho
- {\mcc{R}_{\rho\alpha\beta}}^\rho
+ 2 h^{\rho\sigma} C_{\rho\alpha\sigma\beta}
\notag\\
& - g_{\alpha\beta} G_{\rho\sigma} h^{\rho\sigma}
+ {h_\beta}^\rho G_{\alpha\rho}
+ {h^\rho}_\alpha G_{\rho\beta}
\notag\\
& -(h_{\alpha\beta}- h g_{\alpha\beta}) G/3
+ 4 D^{\rm mG}_{\alpha\beta\gamma\delta} h^{\gamma\delta},
\label{eq:M0}
\end{align}
and \(\mcc{R}_{\alpha\beta\mu\nu} \doteq
\ee^{\ii\theta} (\pd_\nu \pd_\mu \hamp_{\alpha\beta}
+ 2 \ii k_{(\mu} \pd_{\nu)} \hamp_{\alpha\beta}
+ \ii \hamp_{\alpha\beta} \pd_\mu k_\nu)\). 

Note that \m{D^{\text{mG}}_{\alpha\beta\gamma\delta}} and \m{D^{\text{mGEM}}_{\alpha\beta\gamma\delta}} do not have any particular symmetries in the general case. Hence, the symmetry considerations that are commonly used to study waves in vacuum are not necessarily applicable in the presence of plasma. This means that, to find the wave polarization, in the general case one actually has to solve \cref{eq:euler1EM}. In particular, this requires calculating \m{M_{\alpha\beta}}, which can also be written as follows:
\begin{gather}
M_{\alpha\beta} = 4 D^{\text{mG}}_{\alpha\beta\gamma\delta} h^{\gamma\delta} + \mc{O}(\epsilon \hmag).
\label{eq:M1}
\end{gather}
The term \(\mc{O}(\epsilon \hmag)\), which contains the derivatives of the amplitudes, is not necessarily small compared with \(D^{\text{mG}}_{\alpha\beta\gamma\delta} h^{\gamma\delta}\), which is \(\mc{O}(\rho \hmag)\) in the general case \eqref{eq:scaleEM}. Hence, one must either give up the GO approximation and consider the amplitude derivatives along with \(D^{\text{mG}}_{\alpha\beta\gamma\delta} h^{\gamma\delta}\), or neglect the gravitational coupling to the matter, \(M_{\alpha\beta}\), altogether. However, as already noted in \citep{my:gwjeans}, \(M_{\alpha\beta}\) can be retained in special cases where additional large parameters are present. In particular, one such case is the quasistatic limit, where waves have a large refraction index. This limit is discussed in detail in \cref{sec:quasi}. 

Like \cref{eq:waveA1}, equation~\eqref{eq:euler1EM} is invariant with respect to EM gauge transformations \eqref{eq:gaugeEM}, as one can verify by a straightforward calculation using \cref{eq:kaDmGEM}. To address invariance with respect to gravitational gauge transformations \eqref{eq:gaugeG}, one can use the corresponding argument from \citep{my:gwjeans} and adapt it to the case when EM interactions are present. Specifically, let us decompose \cref{eq:euler1EM} into the  longitudinal part (cf. Eq.~(4.20) from \citep{my:gwjeans})
\begin{gather}
k^\alpha M_{\alpha\beta} + 4 k^\alpha a^\mu D^{\rm mGEM}_{\mu\alpha\beta}  = 0
\label{eq:eulerlongEM0}
\end{gather}
and the transverse part (cf. Eq.~(4.21) from \citep{my:gwjeans})
\begin{multline}
\Pi^{\gamma\rho} \Pi^{\delta\sigma}
\big(
k^2 h_{\rho\sigma}
- \bar{M}_{\rho\sigma}
\\
- 4 a^\mu D^{\rm mGEM}_{\mu\rho\sigma} + 2 g_{\rho\sigma} a^\mu D^{\rm mGEM}_{\mu\alpha\beta} g^{\alpha\beta}
\big)
= 0,
\label{eq:eulertransEM0}
\end{multline}
where \m{\Pi^{\alpha\beta} \doteq g^{\alpha\beta} - k^\alpha k^\beta/k^2} is a projection tensor,\footnote{Here, we assume \(k^2 \ne 0\), which is a valid assumption in the presence of plasma with nonzero density. The vacuum case can be considered within this approach as the limit in which the plasma density is nonzero but vanishingly small.} and the overbar represents the trace-reverse of the corresponding rank-2 tensor. According to \cref{eq:eulertransEM0}, one has
\begin{multline}
k^2 h_{\rho\sigma}
- \bar{M}_{\rho\sigma}
- 4 a^\mu D^{\rm mGEM}_{\mu\rho\sigma} + 2 g_{\rho\sigma} a^\mu D^{\rm mGEM}_{\mu\alpha\beta} g^{\alpha\beta}
\\
= \lambda_\rho k_\sigma + k_\rho \lambda_\sigma,
\label{eq:eulertransEM1}
\end{multline}
where \m{\lambda_\alpha} is some vector field. Comparing the trace-reverse of \cref{eq:euler1EM},
\begin{multline}
- k^2 h_{\alpha\beta}
+ k_\alpha k_\gamma {\bar{h}_\beta}^\gamma
+ k_\beta  k_\gamma {\bar{h}_\alpha}^\gamma
+ \bar{M}_{\alpha\beta}
\\
+ 4 a^\mu D^{\rm mGEM}_{\mu\alpha\beta}
- 2 g_{\alpha\beta} a^\mu D^{\rm mGEM}_{\mu\gamma\delta} g^{\gamma\delta} = 0
\label{eq:euler1EMtr}
\end{multline}
one immediately finds from \cref{eq:eulertransEM1} that
\begin{gather}
\lambda_\alpha = k^\beta \bar{h}_{\alpha\beta},
\end{gather}
which are the degrees of freedom always afforded by the metric gauge invariance \cite[Sec.~8.3]{book:schutz}. Thus, up to gauge freedom, the transverse part encodes all the physical information required to determine the solution for the metric perturbation. Then, \cref{eq:eulerlongEM0} can be used as a test of whether the model for the dispersion function \(D^{\rm mG}_{\alpha\beta\gamma\delta}\) preserves gauge invariance. Using \cref{eq:kbDmGEM}, this equation can be further simplified down to
\begin{gather}
k^\alpha M_{\alpha\beta} = 0,
\label{eq:eulerlongEM1}
\end{gather}
which coincides with the corresponding equation for neutral matter in \citep{my:gwjeans}. As shown in \citep{my:gwjeans}, \cref{eq:eulerlongEM1} is also equivalent to
\begin{gather}
M_{\rho\sigma}[2 \Lambda_{(\alpha} k_{\beta)} \ee^{\ii\theta}] = 0,
\end{gather}
where the square brackets denote the argument on which \m{M_{\rho\sigma} [h_{\alpha\beta}]} is evaluated, and \(\Lambda_\alpha\) is an arbitrary small field. The gauge invariance of \cref{eq:waveA1,eq:euler1EM} ensures that the system is not overspecified for the six independent gauge-invariant variables for gravitational modes and the three such variables for EM modes.

\section{Quasistatic limit}
\label{sec:quasi}

\subsection{Basic equations}

As mentioned in \cref{sec:waveeqh}, gravitational coupling with matter can be retained within the GO approximation if there is a large parameter that allows one to ignore the amplitude derivatives in the \(\mc{O}(\epsilon \hmag)\) term in \cref{eq:M1} with respect to the matter coupling term given by \(D^{\text{mG}}_{\alpha\beta\gamma\delta} h^{\gamma\delta}\). A possible candidate for such a parameter is the refraction index $N$. In what follows, we assume the parametrization \(k^\alpha = (\omega, 0, 0, \spatialk)\), so $\omega$ is the wave frequency and \m{\spatialk} is the spatial wavenumber; then \(N \doteq \spatialk/\omega\). The dispersion functions \eqref{eq:DGEMreso} scale as \(\mc{O}(\epsilon^2 N^2)\), as can be seen by expanding the derivatives in \autoref{eq:DGEM}. Thus for a large enough \(N\ \gg 1\), the GO limit can be employed as follows. In this limit, the dispersion functions can be approximated as
\begin{subequations}
\begin{align}
D_{\alpha\beta\gamma\delta}^{\rm mG}
& = \sum_s \int \frac{\dd \boldsymbol{p}}{p^0} \, f_s(x,\boldsymbol{p})\,
\frac{k_\mu}{4}\frac{\pd}{\pd p_\mu}
\left(
\frac{\mc{T}_{\alpha\beta} \mc{T}_{\gamma\delta}}{\Omega}
\right),
\label{eq:DmGGO}
\\[1ex]
D_{\alpha\beta\gamma}^{\rm mGEM}
& = \sum_s \int \frac{\dd \boldsymbol{p}}{p^0} \, f_s(x,\boldsymbol{p}) e_s\,
\frac{k_\mu}{2}\frac{\pd}{\pd p_\mu}
\left(
\frac{\mc{T}_{\alpha\beta} p_\gamma}{\Omega}
\right),
\label{eq:DmGEMGO}
\\[1ex]
D_{\alpha\beta}^{\rm mEM}
& = \sum_s \int \frac{\dd \boldsymbol{p}}{p^0} \, f_s(x,\boldsymbol{p}) e^2_s
k_\mu\,\frac{\pd}{\pd p_\mu}
\bigg(
\frac{\mc{T}_{\alpha\beta}}{\Omega}
\bigg),
\end{align}
\label{eq:DGEMGO}%
\end{subequations}
where \autoref{eq:DGEM} are used, the subdominant terms \(\mc{O}(N^0)\) are ignored, and the remaining dominant terms are \(\mc{O}(N^2)\).\footnote{The smaller terms \(\mc{O}(N^1)\) are also retained in general.} Furthermore, \cref{eq:M0,eq:M1} can be written~as
\begin{gather}
M_{\alpha\beta} = 4 D^{\text{mG}}_{\alpha\beta\gamma\delta} h^{\gamma\delta}.
\label{eq:MGO}
\end{gather}
Then, it is readily seen that \cref{eq:kDGEM,eq:eulerlongEM1} are satisfied identically up to term \(\mc{O}(N^0 \spatialk)\) that are negligible within the assumed approximation. This makes \autoref{eq:DGEMGO} a satisfactory model that retains gauge invariance both with respect to the EM gauge as well as the coordinate gauge. The modes that satisfy this approximation can be called \textit{gravito-electrostatic} by analogy with electrostatic waves such as Langmuir mode and the gravitostatic waves \cite{my:gwjeans} such as the Jeans mode. Similarly, \autoref{eq:DGEMreso} for the dispersion functions can be approximated as
\begin{subequations}
\begin{align}
D_{\alpha\beta\gamma\delta}^{\rm mG}
& = \sum_s \int  \frac{\dd\boldsymbol{p}}{4(p^0)^2}
\frac{\boldsymbol{k}\cdot \pd_{\boldsymbol{p}} f_s}{\omega - \boldsymbol{k} \cdot \boldsymbol{v}}\,
\mc{T}_{\alpha\beta} \mc{T}_{\gamma\delta},
\\[1ex]
D_{\alpha\beta\gamma}^{\rm mGEM}
& = \sum_s \int  \frac{\dd\boldsymbol{p}}{2(p^0)^2}
\, e_s
\frac{\boldsymbol{k}\cdot \pd_{\boldsymbol{p}} f_s}{\omega - \boldsymbol{k} \cdot \boldsymbol{v}}\,
\mc{T}_{\alpha\beta} p_\gamma,
\\[1ex]
D_{\alpha\beta}^{\rm mEM}
& = \sum_s \int  \frac{\dd\boldsymbol{p}}{(p^0)^2}
\, e^2_s
\frac{\boldsymbol{k}\cdot \pd_{\boldsymbol{p}} f_s}{\omega - \boldsymbol{k} \cdot \boldsymbol{v}}\,
\mc{T}_{\alpha\beta}.
\end{align}
\label{eq:DGEMresoGO}%
\end{subequations}
These are equivalent to \autoref{eq:DGEMGO} up to subdominant terms \(\mc{O}(N^0)\), as can be seen by comparing \cref{eq:DGEM,eq:DGEMreso}.

\subsection{Newtonian limit}

Now, let us consider the Newtonian limit,\footnote{Some effects of Newtonian gravity on plasma dynamics have been addressed in the literature; see, for example, \cite{ref:coppi23, ref:pandey99, ref:lundin08, ref:prajapati12}.} in which \(N \to \infty\) and the background energy--momentum tensor is nonrelativistic,
\begin{gather}
\frac{\mc{T}_{\alpha\beta}}{(p^0)^2} \simeq \delta^0_\alpha \delta^0_\beta.
\label{eq:newtonp}
\end{gather}
(This approximation does not exclude all thermal effects; see below.) Pursuing an approach similar to the one used in \citep{my:gwjeans}, let us change the normalization of the distribution functions \(f_s(x, \boldsymbol{p})\,\dd\boldsymbol{p} \to N_s(x) \,f_s(\boldsymbol{v})\,\dd\boldsymbol{v}\), as common in nonrelativistic plasma theory \cite{book:stix}, with $N_s = \rho_s/m_s$. Then, \autoref{eq:DGEMresoGO} can be written as
\begin{subequations}
\begin{align}
D_{\alpha\beta\gamma\delta}^{\rm mG}
& = \frac{1}{4}\sum_s \mc{X}_s
\delta^0_\alpha \delta^0_\beta \delta^0_\gamma \delta^0_\delta,
\\[1ex]
D_{\alpha\beta\gamma}^{\rm mGEM}
& = -\frac{1}{2}\sum_s
\frac{e_s}{m_s}\,\mc{X}_s\delta^0_\alpha \delta^0_\beta \delta^0_\gamma,
\label{eq:DmGEMnewt}
\\[1ex]
D_{\alpha\beta}^{\rm mEM}
& = \sum_s
\frac{e_s^2}{m_s^2}\,\mc{X}_s \delta^0_\alpha \delta^0_\beta,
\end{align}
\label{eq:DGEMresonewt}%
\end{subequations}
where we introduced
\begin{gather}
\mc{X}_s
\doteq \rho_s(x) \int_{\mc{L}}
\frac{\boldsymbol{k}\cdot \pd_{\boldsymbol{v}} f_s(\boldsymbol{v})}{\omega - \boldsymbol{k} \cdot \boldsymbol{v}}\,
\dd \boldsymbol{v}.
\label{eq:XEM}
\end{gather}
Then, \cref{eq:MGO} can be used to write
\begin{subequations}
\begin{gather}
M_{\alpha\beta} = \sum_s  \mc{X}_s\,
\delta^0_\alpha \delta^0_\beta h^{00},
\end{gather}
whence
\begin{align}
\bar{M}_{\alpha\beta}
& = \sum_s  \mc{X}_s h^{00}
\left(
\delta^0_\alpha \delta^0_\beta + \frac{1}{2}\, \eta_{\alpha\beta}
\right)
\notag\\
& = \sum_s  \frac{1}{2} \,\mc{X}_s h^{00} I_{\alpha\beta},
\end{align}
\end{subequations}
where \(I_{\alpha\beta}\) is a unit matrix. Using the EM and gravitational gauge invariance, let us adopt the Lorenz gauge both for the EM potential and the metric perturbation, \ie
\begin{gather}\label{eq:lgauge}
k^\alpha a_\alpha = 0,
\qquad
k^\beta \bar{h}_{\alpha\beta} = 0.
\end{gather}
Then, \cref{eq:waveA1,eq:euler1EMtr} can be written as follows:
\begin{subequations}
\begin{gather}
- k^2 a_\alpha
+ \sum_s \left(
\frac{e_s^2}{m_s^2}\, \mc{X}_s\ a^0
- \frac{e_s}{2m_s}\, \mc{X}_s h^{00}
\right) \delta^0_\alpha
= 0,
\\
- k^2 h_{\alpha\beta}
+ \sum_s  \left(
\frac{1}{2}\, \mc{X}_s h^{00}
- \frac{e_s}{m_s}\, \mc{X}_s\ a^0 
\right) I_{\alpha\beta}
= 0.
\end{gather}
\label{eq:waveeq}%
\end{subequations}
It is seen from here that gravito-electrostatic modes have the longitudinal polarization similar to that found for the Langmuir mode and the Jeans mode, \ie
\begin{gather}
h_{\alpha\beta} = I_{\alpha\beta} \hamp,
\qquad
a_\alpha = \delta^0_\alpha \aamp,
\end{gather}
where \(\hamp\) and \(\aamp\) denote the scalar amplitudes of the respective wave fields. Substituting this into \cref{eq:waveeq} readily yields
\begin{gather}
k^2 \aamp
+ \sum_s \frac{e_s}{m_s} \left(
\frac{1}{2}\, \mc{X}_s \hamp
+ \frac{e_s}{m_s}\, \mc{X}_s \aamp
\right)
= 0,
\\
- k^2 \hamp
+ \sum_s  \left(
\frac{1}{2}\, \mc{X}_s \hamp
+ \frac{e_s}{m_s}\, \mc{X}_s \aamp
\right)
= 0,
\end{gather}
so the wave polarization is found to be
\begin{gather}
\frac{\aamp}{\hamp}
= \frac{k^2 -  W_0/2}{W_1},
\label{eq:polarization}
\end{gather}
where we have introduced the notation
\begin{gather}
W_n \doteq \sum_s  \left(\frac{e_s}{m_s}\right)^n \mc{X}_s.
\end{gather}
The corresponding dispersion relation is given by
\begin{gather}
k^2 = \frac{W_0}{4} - \frac{W_2}{2} \pm \sqrt{\left(
\frac{W_0}{4} + \frac{W_2}{2}\right)^2
-\frac{W_1^2}{2}}.
\label{eq:dispersion}
\end{gather}
Below, we explore this result in several limits.

\subsection{Cold plasma limit}
\label{sec:qscold}

In the cold-plasma limit, where all particle velocities satisfy $v \ll \omega/\msf{k}$ and can be neglected completely, one has
\begin{align}
\int_{\mc{L}} \dd \boldsymbol{v}\,
\frac{\boldsymbol{k}\cdot \pd_{\boldsymbol{v}} f(\boldsymbol{v})}{\omega - \boldsymbol{k} \cdot \boldsymbol{v}}
& \simeq \int \dd \boldsymbol{v}\left(1 + \frac{\boldsymbol{k}\cdot \boldsymbol{v}}{\omega}\right)
\frac{\boldsymbol{k}}{\omega}\cdot \pd_{\boldsymbol{v}} f(\boldsymbol{v})
\nonumber\\
& = \frac{1}{\omega^2} \int \dd \boldsymbol{v}\,
\left( \boldsymbol{k}\cdot \boldsymbol{v}\right) \boldsymbol{k}\cdot \pd_{\boldsymbol{v}} f(\boldsymbol{v})
\nonumber\\
& = -\frac{\msf{k}^2}{\omega^2} \int \dd \boldsymbol{v}\,f(\boldsymbol{v})
\nonumber\\
& = -\frac{\msf{k}^2}{\omega^2}.
\end{align}
This readily yields
\begin{gather}
\mc{X}_s = - \rho_s\, \frac{\spatialk^2}{\omega^2},
\label{eq:chicold}
\end{gather}
Due to the assumed background neutrality \m{\sum_s \left(e_s \rho_s/m_s \right) = 0} [\cref{eq:backj}], this gives $W_1 = 0$. Remember also that \(k^2 \simeq \spatialk^2\), since \(N \gg 1\) is assumed. Then, it is readily seen  that \cref{eq:dispersion} predicts the following two modes. One of them satisfies
\begin{gather}
k^2 = - W_2,
\end{gather}
whence $\omega^2 = \omega_{\rm p}^2$, where $\omega_{\rm p}$ is the plasma frequency:
\begin{gather}
\omega_{\rm p}^2 \doteq \sum_s \frac{e^2_s}{m^2_s}\, \rho_s.
\label{eq:omegal}
\end{gather}
The corresponding polarization is \((\hamp, \aamp) = (0,1) \times \text{const}\). This is the familiar Langmuir mode \cite{book:stix}. The other mode predicted by \cref{eq:dispersion} satisfies
\begin{gather}
k^2 = \frac{W_0}{2}.
\end{gather}
This leads to $\omega^2 = - \omega_{\rm J}^2$, where $\omega_{\rm J}$ is the Jeans frequency:
\begin{gather}
\omega_{\rm J}^2 \doteq \frac{1}{2} \sum_s \rho_s.
\label{eq:omegaj}
\end{gather}
The corresponding polarization is \((\hamp, \aamp) = (1,0) \times \text{const}\). This is the familiar Jeans mode \cite{ref:trigger04, ref:lima02, ref:ershkovich08}. Notice that
\begin{gather}
\omega^2_{\rm p} \simeq e^2 \rho_e/m_e^2,
\qquad
\omega^2_{\rm J} \simeq \rho_i/2,
\label{eq:omegas}
\end{gather}
so \m{\omega^2_{\rm J}/\omega^2_{\rm p} \ll 1} by \cref{eq:scaleEM1}.

As a side remark, note that our derivation of the Jeans mode also presents an alternative resolution of the ``Jeans swindle", the problem that the traditional derivation of the Jeans mode arbitrarily ignores the background gravitational field, which is necessarily present in the presence of matter \cite{book:binney11, tex:ershkovich11}. The proposed resolutions to the Jeans swindle are to assume infinite homogeneous background \cite{tex:ershkovich11} or to cancel out the background effect by the Hubble expansion \cite{ref:falco13}. In contrast, our derivation arrives at the local Jeans dispersion relation using formal asymptotic analysis of the linear GWs even for a Minkowski background with an inhomogeneous mass distribution.

Also note that the absence of coupling between the electrostatic and gravitational modes in the cold-plasma limit is a general statement that is valid also beyond the quasistatic approximation. This is seen from the fact that this interaction is determined by the ``cross term'' \m{\favr{a^{\alpha}, D_{\alpha\beta\gamma}^{\rm mGEM} h^{\beta\gamma}}} in the action \eqref{eq:S2m}. As easily seen from \cref{eq:DmGEM}, the dispersion function \(D_{\alpha\beta\gamma}^{\rm mGEM}\) vanishes in the cold limit due to neutrality of the background plasma. Thus, in the cold limit, gravitational modes and EM modes cannot interact (within the model assumed in this paper) irrespective of the value of \(k^\alpha\). This can be considered as an extension of the results obtained in \citep{ref:flauger18}, where no dispersion of the tensor modes was observed for their propagation through cold neutral matter.

\subsection{Warm plasma}
\label{sec:warm}

In a warm plasma, when the characteristic speeds of particles are small compared to $\omega/\msf{k}$ but not entirely negligible, the coupling term \(W_1\) is small but nonzero and can be treated perturbatively. In this case, the square root in \cref{eq:dispersion} can be Taylor-expanded in $W_1$. This results in small modifications of the Jeans mode and the Langmuir mode. For the modified Jeans mode, one obtains
\begin{subequations}
\begin{gather}
k^2 = 
- \frac{\spatialk^2}{\omega^2}\, \omega_{\rm J}^2
-\frac{W_1^2}{2 W_2},
\label{eq:jeanswarm1}
\\
\frac{\aamp}{\hamp} = - \frac{W_1}{2 W_2},
\end{gather}
\label{eq:jeanswarm}%
\end{subequations}
where we used \cref{eq:omegaj}, and the modified Langmuir mode satisfies
\begin{subequations}
\begin{gather}
k^2 = 
\frac{\spatialk^2}{\omega^2}\, \omega_{\rm p}^2
+ \frac{W_1^2}{2 W_2},
\label{eq:langmuirwarm1}
\\
\frac{\aamp}{\hamp} = 
- \frac{W_2}{W_1}
+ \frac{W_1}{2 W_2},
\end{gather}
\label{eq:langmuirwarm}%
\end{subequations}
where we used \cref{eq:omegal} and \(W_0 \ll W_2\). 

For example, let us consider a specific example of a warm Maxwellian plasma consisting of electrons with temperature $T_e = m_e v_{Te}^2$ and ions with comparable (or smaller) temperature $T_i = m_i v_{Ti}^2$. Due to the smallness of the ratio $m_e/m_i$ and the assumed condition $v_s \ll \omega/\spatialk$ for all $s$, one has $v_{Ti} \ll v_{Te} \ll \omega/\spatialk$. This renders the ion temperature negligible, while the electron temperature can be treated as a small but nonvanishing correction to the case considered in \cref{sec:qscold}. Then, one obtains \cite{book:stix}
\begin{gather}
\mc{X}_e \simeq - \rho_e\,\frac{\spatialk^2}{\omega^2}
\left(
1 + \frac{3 \spatialk^2 v_{Te}^2}{\omega^2}
\right),
\\
\mc{X}_i \simeq - \rho_i\,\frac{\spatialk^2}{\omega^2},
\end{gather}
whence
\begin{gather}
W_1 \simeq 3\,e N_e\, \frac{\spatialk^4 v_{Te}^2}{\omega^4}.
\end{gather}
Using this in \cref{eq:jeanswarm1,eq:langmuirwarm1} immediately leads to the dispersion relation, \(\omega^2 = -\omega^2_{\rm J} + \Delta^2\) for the modified Jeans mode, and \(\omega^2 = \omega^2_{\rm p} - \Delta^2\) for the modified Langmuir mode respectively, with
\begin{gather}
\frac{\Delta^2}{\omega_{\rm J}^2}
\simeq \frac{9m_e}{m_i} \, \frac{\spatialk^4 v_{Te}^4}{\omega^4},
\quad
\frac{\Delta^2}{\omega_{\rm p}^2}
\simeq \frac{9m_e^2}{2 e^2} \, \frac{\spatialk^4 v_{Te}^4}{\omega^4},
\label{eq:aux2}
\end{gather}
where we have retained only the leading-order terms and neglected corrections \(\mc{O}(\spatialk^2 v_s^2/\omega^2)\). The relative strength of the correction for the Jeans mode is very small, which is in line with the cold plasma limit of no interaction at all. Similarly, the relative strength of the modification of the Langmuir mode due to gravitational effects is further suppressed by an additional factor of \(m_i m_e/e^2 \ll 1\), due to the gravitational forces being weaker compared to the EM forces.

\subsection{Plasma with cold ions and hot electrons}

A more significant interaction between the Jeans mode and plasma electrostatic modes can be found at low frequencies that satisfy
\begin{gather}
v_{Ti} \ll \omega/\msf{k} \ll v_{Te}.
\end{gather}
In this regime, \cref{eq:XEM} can be approximated as \cite[Sec.~8-13]{book:stix}
\begin{gather}
\mc{X}_e = \frac{\rho_e}{v_{Te}^2},
\qquad
\mc{X}_i = -\frac{\rho_i \spatialk^2}{\omega^2}.
\end{gather}
In the absence of gravitational interactions, such a plasma supports ion acoustic oscillations
\begin{gather}\label{eq:aux3}
\omega^2 \simeq \frac{k^2 c_{\rm s}^2}{1 + k^2 \lambda_{De}^2} \to k^2 c_{\rm s}^2.
\end{gather}
Here, \m{\lambda_{De} \doteq v_{Te}/\omega_{\rm p}} is the electron Debye length, \(c_{\rm s} \doteq (Z T_e/m_i)^{1/2}\) is the ion sound speed (assuming $e_i = Ze$), and the second part of \cref{eq:aux3} corresponds to the limit \m{k^2 \lambda_{De}^2 \ll 1}, where the dispersion relation becomes particularly simple (sound-like). Although electric charge \(e\) does not enter \cref{eq:aux3} explicitly, the electrostatic interactions are important in that they tie cold ions, which provide inertia, to hot electrons, which carry pressure.

Now, let us reinstate gravitational interactions, assuming \(\spatialk^2 c_{\rm s}^2 \sim \omega^2_{\rm J}\). In this case, the square root in the dispersion relation \eqref{eq:dispersion} can still be Taylor-expanded in \(W_1\) as per \cref{eq:jeanswarm,eq:langmuirwarm}, since
\begin{align}
\frac{W_1}{W_2}
&= \left(-1 - \frac{\spatialk^2 v_{Te}^2}{\omega^2}\right)
\left(\frac{e}{m_e} - \frac{Z e \spatialk^2 v_{Te}^2}{m_i \omega^2}\right)^{-1}
\notag \\
& \simeq 
\frac{m_i}{e}\left(Z - \frac{m_i /m_e}{ \spatialk^2 v_{Te}^2 / \omega^2} \right)^{-1}
\end{align}
is small, as can be verified a posteriori [using \cref{eq:jeansion}]. Then,
\begin{align}
\frac{W_1^2}{W_2}
& \simeq - \rho_e\, \frac{\spatialk^2}{\omega^2}
\left(-\frac{\omega^2}{\spatialk^2 v_{Te}^2} + \frac{Z m_e}{m_i}\right)^{-1}
\notag \\ 
& = \rho_i \,\frac{\spatialk^2}{\omega^2}
\left(\frac{\spatialk^2 c_{\rm s}^2}{\omega^2 - \spatialk^2 c_{\rm s}^2}\right),
\label{eq:correction}
\end{align}
so \cref{eq:jeanswarm1} leads to the following dispersion relation:
\begin{gather}
\omega^2 = 
-\omega_{\rm J}^2 + \spatialk^2 c_{\rm s}^2,
\label{eq:jeansion}
\end{gather}
where we have again used the approximation \(\omega_{\rm J}^2 \simeq \rho_i/2\). One can understand this as the Jeans mode hybridized with the ion-acoustic branch. (For comparison: in a neutral gas, the ion sound speed in the Jeans mode's dispersion relation is replaced by the particle thermal speed \cite{my:gwjeans}.)

One can similarly calculate the correction to the ion acoustic waves that result from their hybridization with the Jeans mode. It is easy to see that this correction is of order  \(\sim m_e m_i/e^2\). Thus, the effect of gravitational interactions on electrostatic waves is negligible, which is again due to the relative weakness of the gravitational forces compared with the EM forces.

\section{Transverse waves}
\label{sec:trans}

Now let us consider transverse waves. To the extent that gravitational effects can be neglected, transverse EM waves in nonrelativistic plasma satisfy the well-known GO dispersion relation
\begin{gather}\label{eq:EMw}
k^2 = -\omega^2_{\rm p},
\end{gather}
and GWs satisfy the GO dispersion relation
\begin{gather}
k^2 = 0,
\end{gather}
because corrections due to the interaction with the matter are beyond the GO approximation \cite{my:gwjeans}. Let us consider the interaction of these waves perturbatively.

Assuming the Lorenz gauge \eqref{eq:lgauge} again, \cref{eq:waveA1,eq:euler1EMtr} can be written as follows:
\begin{gather}
- k^2 a_\alpha
+ D^{\rm mEM}_{\alpha\beta} a^\beta
+ D^{\rm mGEM}_{\alpha\beta\gamma} h^{\beta\gamma}
\label{eq:waveALor}
= 0,
\\
- k^2 h_{\alpha\beta}
+ \bar{M}_{\alpha\beta}
+ 4 a^\mu D^{\rm mGEM}_{\mu\alpha\beta}
- 2 g_{\alpha\beta} a^\mu D^{\rm mGEM}_{\mu\gamma\delta} g^{\gamma\delta} = 0.
\label{eq:wavehLor}
\end{gather}
Let us first consider the mode whose frequency is closest to that given by the EM dispersion relation \eqref{eq:EMw}. For this mode, the last term in \cref{eq:waveALor} is a small perturbation and \(h_{\alpha\beta}\), can be found perturbatively from \cref{eq:wavehLor} by substituting \(k^2 \simeq - \omega^2_{\rm p}\) there, \ie adopting
\begin{gather}
\omega_{\rm p}^2 h_{\alpha\beta}
+ \bar{M}_{\alpha\beta}
+ 4 a^\mu D^{\rm mGEM}_{\mu\alpha\beta}
- 2 g_{\alpha\beta} a^\mu D^{\rm mGEM}_{\mu\gamma\delta} g^{\gamma\delta} = 0.
\end{gather}
Since $N \sim 1$ for such waves, one has \(M_{\alpha\beta} \sim \mc{O}(\epsilon)\), so \(M_{\alpha\beta}\) can be ignored. Let us also simplify our notation by introducing \(D_{\alpha\beta\gamma} \doteq D_{\alpha\beta\gamma}^{\rm mGEM}\). Then, one can write that
\begin{gather}
h_{\alpha\beta}
= - \frac{1}{\omega_{\rm p}^2} \left( 4 a^\mu D_{\mu\alpha\beta}
- 2 g_{\alpha\beta} a^\mu D_{\mu\gamma\delta} g^{\gamma\delta} \right).
\end{gather}
By substituting this into \cref{eq:waveALor}, one obtains
\begin{gather}
- k^2 a_\alpha + \mc{D}^{\rm mEM}_{\alpha\mu} a^\mu = 0,
\end{gather}
where \m{\mc{D}^{\rm mEM}_{\alpha\mu}} can be understood as the EM dispersion function dressed by gravitational interactions:
\begin{gather}
\mc{D}^{\rm mEM}_{\alpha\mu} \doteq D^{\rm mEM}_{\alpha\mu}
- 2\omega_{\rm p}^{-2} D_{\alpha\beta\gamma}
\left(2 {D_{\mu}}^{\beta\gamma}
- g^{\beta\gamma} {D_{\mu\delta}}^\delta \right).
\label{eq:Deff}
\end{gather}

Since \(D_{\alpha\beta\gamma}\) vanishes in the the cold-plasma limit, the difference between \m{\mc{D}^{\rm mEM}_{\alpha\mu}} and \m{D^{\rm mEM}_{\alpha\mu}} is entirely due to finite particle speeds (in the frame specified in \cref{sec:asm}). This difference can be estimated as follows. Assuming nonrelativistic speeds, one has \m{p_\alpha \simeq -m \delta^0_\alpha}, so \m{a^\mu D_{\mu\alpha\beta} \simeq a^0 D_{0\alpha\beta}}. Then, like in \cref{sec:warm}, one finds that
\begin{gather}\label{eq:Dest}
D_{0\alpha\beta} \sim \frac{\spatialk^2 v_{Te}^2}{\omega^2}\,N e,
\end{gather}
and $\spatialk v_{Te} \ll \omega$ in nonrelativistic plasma, because ${\omega/\spatialk \gtrsim 1}$. Also note that \(D^{\rm mEM}_{\alpha\beta} \sim \omega_{\rm p}^2\) and \(\omega^2_{\rm p} \sim N e^2/m_e\). Then, the second term on the right-hand side of \cref{eq:Deff} scales relative to the first term as
\begin{gather}
\left(\frac{\spatialk v_{Te}}{\omega}\right)^4 
\frac{N^2 e^2}{\omega_p^4}
\lesssim 
\frac{m_e^2}{e^2}.
\end{gather}
In other words, the frequency shift of an EM waves in plasma due to gravitational effects is of order \m{\mc{O}(m_e^2/e_e^2)}, which is extremely small. Note that this conclusion holds even in the limit of vanishingly small plasma density, where GWs and EM waves have the same dispersion relation, $k^2 = 0$. This is due to the fact that both the deviation from the linear resonance between these waves and their coupling ($\sim D^2/\omega_{\rm p}^2$) are equally proportional to $N$, so these waves remain nonresonant even at $N \to 0$.

A similar procedure can be used to obtain the correction to the dispersion relation for the GWs with $a_\alpha$ treated as a perturbation. Substituting \(k^2 = 0\) in \cref{eq:waveALor} yields
\begin{gather}
D^{\rm mEM}_{\alpha\beta} a^\beta
= - D_{\alpha\beta\gamma} h^{\beta\gamma},
\\
D_{\alpha\beta}^{\rm mEM}
= -\omega^2_{\rm p} g_{\alpha\beta} + \sum_s e^2_s \int \frac{\dd \boldsymbol{p}}{p^0} \, f_s(x,\boldsymbol{p}) \,
\frac{k_\alpha p_\beta}{k^\mu p_\mu},
\end{gather}
where \cref{eq:DmEM} is used to obtain the expression for \(D_{\alpha\beta}^{\rm mEM}\). Let us denote the inverse of \m{D^{\rm mEM}_{\alpha\beta}} as \m{D_{-1}^{\alpha\beta}}. Then,
\begin{gather}
a^\rho = - D_{-1}^{\rho\alpha} D_{\alpha\beta\gamma} h^{\beta\gamma}.
\end{gather}
By substituting this to \cref{eq:wavehLor}, one obtains
\begin{multline}
- k^2 h_{\alpha\beta}
+ \bar{M}_{\alpha\beta}
- 4 D_{-1}^{\mu\rho} D_{\rho\sigma\tau} h^{\sigma\tau} D_{\mu\alpha\beta}
\\
+ 2 g_{\alpha\beta} D_{-1}^{\mu\rho} D_{\rho\sigma\tau} h^{\sigma\tau} {D_{\mu\gamma}}^\gamma = 0.
\label{eq:htrans}
\end{multline}
The last two terms in \cref{eq:htrans} represent the correction caused by GW coupling with EM field. Like in the case with EM waves, let us use that \m{D_{\alpha\beta} \sim \omega_{\rm p}^2} and \m{D_{\alpha\beta\gamma} \ll Ne}. Then, said correction is much less than~\m{\rho_e}, which is negligible compared with \m{M_{\alpha\beta} \sim \rho_i} [see \cref{eq:chicold}, with \m{\spatialk^2 \simeq \omega^2}]. This means that EM effects in nonrelativistic nonmagnetized have a much smaller (at least by the factor $m_e/m_i \ll 1$) effect on GWs than gravitational interactions with the plasma. Furthermore, as we have pointed out earlier, even the effect of the latter is negligible within the GO approximation. In this sense, our results agree with the existing literature in that, \textit{within the GO limit}, there is no linear coupling between the gravitational modes and the EM modes in any smooth background. The modifications of the dispersion relations of tensor gravitational modes that have been earlier reported for specific backgrounds \cite{ref:chesters73,ref:madore73, ref:madore72, ref:barta18,ref:flauger18, ref:baym17, ref:asseo76} are beyond GO and therefore are not in disagreement with our findings.

\section{Conclusions}
\label{sec:conc}

In summary, here we explore the hybridization of linear metric perturbations with their electromagnetic (EM) counterpart in non-magnetized plasma without assuming any special symmetry of the background. We begin, in \cref{sec:plasmaem}, with the derivation of the effective (``oscillation-center'') Hamiltonian \eqref{eq:S2m} that governs the average dynamics of plasma particles in a prescribed quasimonochromatic wave that involves metric perturbations and EM fields simultaneously. Then, using this Hamiltonian, in \cref{sec:waveeqEM}, we derive the backreaction of plasma particles on the wave itself and obtain the equations [\eqref{eq:waveA1} and \eqref{eq:euler1EM}] that describe the resulting self-consistent gravito-electromagnetic (GEM) waves in a plasma. We also demonstrate the gauge invariance of these equations and discuss how this coordinate invariance can be used to test \eqref{eq:kDGEM} other approximate models of gravitational-wave (GW) dispersion in plasma.

We find that in the cold plasma limit (\cref{sec:qscold}), linear gravitational modes and EM modes do not interact within the model assumed here, necessitating thermal effects for any such interaction. In a sufficiently dense plasma, \textit{transverse} GEM modes consist of modes similar to the familiar transverse EM waves in plasma and the tensor modes of the GWs in vacuum, respectively. The shift of the GW frequency due to plasma is found to be generally of the same order as diffraction caused by plasma's curving the background spacetime; therefore, it is beyond the accuracy of the geometrical-optics (GO) approximation, and the transverse tensor modes have no significant interaction with the plasma and the EM modes in the GO limit. However, for \textit{longitudinal} GEM modes with large values of the refraction index, the interplay between gravitational and EM interactions in plasma can have a strong effect.  In the case of purely gravitational interactions, these longitudinal modes contain the so-called scalar modes that also yield the Newtonian Jeans mode in the limit of large refraction index. In particular, the dispersion relation of the Jeans mode is significantly affected by electrostatic interactions. The approach used in this work can also be readily extended to magnetized plasma, an endeavour that we leave to future work.

This material is based upon the work supported by National Science Foundation under the grant No.\ PHY~1903130.

\bibliography{main,my,gw,cosmo}

\end{document}